\documentclass[aps,twocolumn,showpacs]{revtex4} 
\usepackage{graphics,epsfig}

\newcommand{\be}{\begin{equation}}
\newcommand{\ee}{\end{equation}}
\newcommand{\ba}{\begin{eqnarray}}
\newcommand{\ea}{\end{eqnarray}}
\newcommand{\lb}{\label}
\newcommand{\bb}{\bibitem}
\newcommand{\half}{\frac{1}{2}}

\newcommand{\nn}{\nonumber}
\begin{document}

\title{On the Stability of Circular Orbits of Particles Moving around  Black Holes Surrounded by Axially Symmetric Structures}

\author{ Patricio.S. Letelier\footnote{e-mail: letelier@ime.unicamp.br}
 } 
 
\affiliation{
 Departamento de Matem\'atica Aplicada-IMECC,
Universidade Estadual de Campinas,
13081-970 Campinas,  S.P., Brazil}

\begin{abstract}

The Rayleigh criterion  is used 
 to    study the stability of circular orbits   of particles moving 
around static black  holes surrounded by different axially symmetric 
 structures with reflection symmetry,  like disks, rings and halos.
We consider three  models of disks one of infinite extension and  two
finite,  and one model of rings. The halos are represented by external
 quadrupole moments (either oblate  or  prolate). Internal  quadrupole
 perturbation (oblate and prolate) are also considered. For this 
class of  disks the counter-rotation hypothesis 
implies that the  stability of the disks is  equivalent to the stability of
 test particles. The stability of  Newtonian systems is also considered and compared with the equivalent relativistic situation. We find that 
the general relativistic dynamics favors the formation of rings.

\end{abstract}
\maketitle
\setlength{\parindent}{3em}

\section{Introduction}
One of the main tools used to study  the gravitational field of any
 distribution of mass, e.g., a black hole (BH), 
is to examine the motion   of nearby  test particles. In other words, the
complete  understanding of the
different   solutions of the geodesic equation leads us to the  
comprehension of the gravitational properties of the center 
of attraction, in the present case  a black hole surrounded by and axially 
symmetric structure.
Because of its mathematical simplicity and physical relevance   
 circular orbits play a particularly important role in the study of 
the gravitational properties of  a central body. Also circular orbits  usually 
represent   limit situations between different regimes.

The purpose of this paper is to study the stability of circular 
orbits in different gravitating systems formed by a BH  surrounded by
 axially symmetric structures. In particular, we study orbits in BH + disks systems, BH + rings systems, and BH  +multipolar fields. In the case of orbits
 inside  the matter, e.g. a disk, we have that in general the stability 
of the circular orbits can be associated to the stability of the matter itself.
The method used to study the stability  is the
 Rayleigh criteria of stability (RCS) for a  rotating 
 fluid \cite{Lord} that can be easily adapted to the study of  orbits of test particles.  In the context of  general relativity the RCS  has been 
used  to study  the stability of different classes of self-gravitating 
 disks, see for instance  \cite{LB}-\cite{voghtletdisk}. 

The paper is  divided as follows:
In Sec. II we review the RCS and in Sec. III the Einstein equations for the superposition of a static BH and an axially symmetric structure with reflection symmetry is presented.
In Sec.  IV   the stability of circular orbits moving around 
the superposition of a BH and  three models of disks are considered, one of infinite extension and  two finite. In Sec. V we study the stability 
of particles orbiting on the plane of a   ring + BH system,  inside as well as
 outside the ring. 
In Sec. VI we consider circular orbits for a BH perturbed  by either
 an internal
quadrupole field  or a external quadrupole field. We consider 
 oblate as well as prolate quadruples. We have that some general features
 found in the study 
of the previous examples can be simulated with these quadrupole fields. 
 In the last section we discuss some of the previously found results.
We also review  in an Appendix the so called counter-rotating hypothesis.

\section{Rayleigh criteria of stability }

 Consider, in Newtonian dynamics,  a small body of mass $\mu$ moving 
in a circular 
orbit of radius $r=r_0$  around a fix center of gravitational attraction. 
In the non inertial frame
associated to the moving body we have an equilibrium situation: the
 gravitational
 force [$F_g(r_0)$] is equal to the centrifugal force [$F_c(r_0)=L^2(r_0)/(\mu r_{0}^3)],$ where
$L$ is the angular momentum of the particle of mass $\mu$. Now let
 us virtually
 displace  the moving  particle to a higher orbit  ($r>r_0$) keeping 
the same angular
 momentum. Then the centrifugal force in this new position
 is $\hat{F}_c(r)=L^2(r_0)/(\mu r^3)]$. To have a situation of
 equilibrium the displaced 
particle should  move downward, i.e., in direction to  the initial orbit.
 In other words,
 the gravitational force in $r$ should be greater
 than the centrifugal force, $\hat{F}_c(r) $,
 i.e.,    $F_g(r)>\hat{F}_c(r)$, but  $F_g(r)=F_c(r)=L^2(r)/(\mu r^3)$. Hence $L^2(r)>L^2(r_0)$,
  by 
doing a Taylor expansion of $L^2(r)$ around $r=r_0$ we find that, $L\frac{dL}{dr}>0$,
 for a stable  circular orbit.  Rayleigh  in 
 the original derivation of his
 criterion \cite{Lord},    considers  a  rotating ring of fluid  under the force of   a pressure gradient 
instead of a   the particle moving
 in a circular orbit attracted by a gravitational field.

 Now let us consider a particle moving in a static background metric
 with axial symmetry in spherical coordinates, $(t, R, \vartheta, \varphi)$,
\be
 ds^2=g_{tt}dt^2 + g_{RR}dR^2 +  g_{\vartheta \vartheta }d\vartheta^2
 +g_{\varphi\varphi}d\varphi^2, \label{metsph}  
\ee
where the metric functions $g_{\mu\nu}$ depends only on the variables $R, \vartheta$.
The geodesic equation for a circular motion on the plane $\vartheta=\pi/2$, gives us the motion equation,
\be
g_{tt,R}\dot{t}^2+g_{\varphi\varphi,R}\dot{\varphi}^2=0, \label{geo1}  
\ee
and  the constants of motion,
\ba
&&1=g_{tt}\dot{t}^2+g_{\varphi\varphi}\dot{\varphi}^2, \label{const1}    \\
&&E=g_{tt}\dot{t},  \label{const2}  \\
&&h= g_{\varphi\varphi}\dot{\varphi},\label{const3}  
\ea
where the orverdot denotes derivative with respect to the proper time $s$
and  $g_{tt,R}=\partial_R g_{tt}$, etc. The constant $E$ represents the relativistic specific energy and $h$ the specific angular momentum.

Note that the motion equation (\ref{geo1}) can be cast as a balance equation valid on the plane $\vartheta=\pi/2$ ,
\be
g_{tt,R}E^2/g_{tt}^2 = -g_{\varphi\varphi,R}h^2/g_{\varphi\varphi}^2.
\label{bal}  
\ee
So as in the Newtonian case we have a balance between the ``gravitational force''
 and the ``centrifugal force''. Thus, assuming the metric represents the gravitation of a  central body  we will have stability of  circular orbits on the plane  $\vartheta=\pi/2$, when
\be
hh_{,R}>0.
  \label{RCS}
\ee
From (\ref{geo1})- (\ref{const3})we find,
\ba
&&h^2= -\frac{g_{\varphi\varphi}^2   g_{tt,R} }{  g_{tt} g_{\varphi\varphi,R} -
 g_{tt,R} g_{\varphi\varphi}  }, \label{hh}  \\
&&E^2=\frac{  g_{tt}^2 g_{\varphi\varphi,R}}{  g_{tt} g_{\varphi\varphi,R} -
 g_{tt,R} g_{\varphi\varphi}  }. \label{EE}  
\ea

Since gravity is an attractive force we can repeat the same reasoning that leads to (\ref{RCS}),  but now  with  the constant $E(R,\pi/2)$ instead of $h(R,\pi/2)$, we find
\be
EE_{,R}>0.
  \label{RCS2}
\ee
From (\ref{hh}) and (\ref{EE}) we get,
\be
hh_{,R} =-\frac{g_{\varphi\varphi}  }{  g_{tt} }EE_{,R}.  \label{consc}  
\ee
But, always $-g_{\varphi\varphi}/ g_{tt}>0 $, then the  relation
 (\ref{RCS2}) gives us the same information than (\ref{RCS}), in other words we have consistence.

For the stability of circular orbits  in stationary axisymmetric spacetimes 
see Ref. \cite{bardeen} and for stability of rotating fluids in a relativistic stars see Ref.  \cite{seguin}. This last reference presents a very general
 study of the stability of 
 relativistic fluids  that generalizes the  Lord Rayleigh work by considering
  a  relativistic  fluid with both bulk and dynamical viscosity and also 
heath flow. A  slightly different  treatment of the stability of circular orbits like the one presented here can be found in Ref. \cite{marek}.

\section{Weyl metrics: black holes + axially symmetric structures}

 The external gravitational field  produced by an axially symmetric
body can be well  described by the Weyl metric  \cite{W},
\be
ds^2=e^{2\psi}dt^2 -e^{-2\psi}[r^2 d\varphi^2+e^{2\gamma}(dr^2+dz^2)],
 \lb{weylm}
\ee
where the functions $\psi$ and $\gamma$ depend only on the cylindrical coordinates $r$ and $z$.  The vacuum Einstein equations ($
R_{\mu \nu}=0 $) reduce to the usual Laplace equation in
 cylindrical coordinates,
\be
\psi_{,rr}+\psi_{,r}/r+\psi_{,zz}=0, \lb{lapw} 
\ee
and the quadrature,
\be 
d\gamma[\psi]= r[(\psi_{,r}^2-\psi_{,z}^2)dr+2\psi_{,r}\psi_{,z}dz].
 \lb{gawc}
\ee
When  $\psi$ satisfies the Laplace equation this differential is exact. 

   The Schwarzschild solution in Weyl coordinates takes the form,
 \ba
 \psi_S&=&\frac{1}{2}\ln\frac{R_++R_- -2m}{R_++R_-+2m},  \lb{psisw}\\
\gamma_S&=&\half\ln\frac{(R_++R_-)^2-4m^2}{R_+R_-}, \lb{gasw}
\ea
where
\be
R_\pm=\sqrt{r^2+(z\pm m)^2}. \lb{Rpm}
\ee
The function $\psi_S$ is just the Newtonian potential of a  bar  of
 length $2m$ and density $1/2$. The  Weyl coordinates has the drawback that
 the Schwarzschild metric does not look
spherically symmetric and that the horizon is squeezed into a line of 
length $2m$. But has the advantage that one of the Einstein equations is the Laplace equation that is linear, hence it allows to consider different composite systems like a black hole surrounded by  disks or halos in an exact way.

The coordinate transformations,
\be
  r =[R(R-2m))]^\half \sin\vartheta ,\;\; z=(R-m)\cos\vartheta, \lb{coort}
\ee
takes  the black hole metric (\ref{weylm}) with (\ref{psisw})-(\ref{gasw})
 into the  usual Schwarzschild  form.

We  shall consider solutions of the form,
\be
\psi=\psi_S +\hat{\psi}, \lb{supp}
\ee
where $\hat{\psi}(r,z)$ is a
  solution of either Laplace equation  or Poisson equation with a density that has support on the plane $z=0\;\; ( \vartheta=\pi/2)$.
The function $\gamma$ in this case is also   written as
\be
\gamma[\psi]=\gamma_S+\hat\gamma(r,z).  \lb{supg}
\ee
In the case of vacuum solutions $\hat\gamma(r,z)$ can be computed directly from
(\ref{gawc} ) and (\ref{supp}).

Therefore  for  the superposition (\ref{supp})
the Weyl metric takes the form,
\ba
ds^2=(1-\frac{2m}{R})e^{2\hat\psi}dt^2-
\frac{e^{2(\hat\gamma-\hat\psi)}}{1-\frac{2m}{R}}dR^2 \nonumber \\
 -R^2 e^{-2\hat\psi}(e^{2\hat{\gamma}}d\vartheta^2 +\sin^2\vartheta d\varphi^2  ). \lb{sp}
\ea
When $\hat\psi =\hat\gamma=0$ we recover the Schwarzschild metric.

We shall analyze three  types on superpositions of BH and matter. In the
 first, 
the matter is represented by a potential 
$\hat\psi$ that describes a thin disk. In 
this case  we will have   that  only     two components of
 energy-momentum tensor (EMT),
 $T_{t}^{t}$ and $T_{\varphi}^{\varphi}$, are   different from zero  
 on the plane of the disk. Thus we have a disk with no radial pressure or
 tension to support the gravitational
 attraction. Since the Weyl metric is static we  have no 
 rotation. Then a counter-rotation hypothesis is needed to have a
 stable situation: we assume  that on the disk as many particles
 move clockwise as counterclockwise (see Appendix A). Even though, this
 interpretation can be 
seen as a device, there is  observational evidence of disks 
made of streams of  rotating and counter-rotating matter \cite{counter}. We
 believe that
counter rotating hypothesis appeared first, in this context, in Ref. \cite{mm}.

 In the general case we will have radial pressure and magnetic  forces
and our simpler analysis need to be modified. In the case when no magnetic
 fields are present and the 
density and pressure of the matter of the disk  are low, e.g., a 
diluted gas of stars, the main contribution to the force on each fluid
 particle is the gravitational force that is taken into account  through the 
spacetime metric.

In the second case we study the superposition of a BH and a thin ring. 
We consider orbits inside the ring as well as orbits outside the ring. And
in the  third case  we consider the superposition of a BH  with multipolar fields. We first consider internal multipolar perturbations, 
i.e., $\hat\psi$ is  a   multipolar field that is zero at spatial infinite.
 In this way  we can describe the gravity outside 
 compact systems formed by 
 a true black hole (or a dense object)
 surrounded by  a distribution of matter like a ring or a small disk. Also we can have an  axially symmetric static dense object with either polar deformations or polar jets. 
We also consider the   case of  a BH with external  multipolar  perturbations, 
i.e., $\hat\psi$ is  a   multipolar field, solution of Laplace equation
 (\ref{lapw}), that diverges  at spatial infinite.  
 In this case we can represent the gravity in the region
 limited by a black hole
and an outer axially symmetric shell of matter or halo.

All the models of BH  with surrounded matter
 that we consider in this paper are
constructed from superposition of the Weyl solutions of the form
 (\ref{supg}) and its associated spacetime is described by the metric 
  (\ref{sp}). In this case the constant of motion  (\ref{const3}) takes the form,
\be 
h^2= \frac{R^2e^{-2\hat \psi}[m+R(R-2m)\hat\psi_{,R}]}{R-3m -2(R-2m)\hat\psi_{,R}},
\label{hhsp} 
\ee
and  its derivative can be cast us,
\be
hh_{,R}=  \frac{Re^{-2\hat \psi}[m(R-6m) +F_1]}{2(R-3m-2F_2)},\label{hhrsp} 
\ee
where
\ba
F_1&=&R(28m^2-18mR+3R^2)\hat\psi_{,R}- \nonumber\\
   &&  6R^2(R-3m)(R-2m)\hat\psi_{,R}^2 +4R^3(R-2m)^2\hat\psi_{,R}^3\nonumber\\
  && +R^2(R-m)(R-2m)\hat\psi_{,RR},\label{F1}\\
F_2&=&R(R-2m)\hat\psi_{,R}\label{F2}
\ea
In the present case the motion is limited to the plane $z=0 (\vartheta=\pi/2)$.
From (\ref{coort}) we have that  $r =[R(R-2m))]^\half$ on the plane and 
$ \hat\psi(r,0)= \hat\psi( [R(R-2m))]^\half,0)$. Note that when
 $\hat\psi=0$, pure BH case, from (\ref{hhrsp}) and  $hh_{,R}=0$ we have
 $R=6m$, as required. This orbit is known as the last stable circular orbit (LSCO) or the marginally stable  circular orbit (MSCO). In our study of stability we will use Schwarzschild  like
 coordinates. The use of Weyl coordinates or proper distance 
coordinates are known to give  similar
  qualitative results \cite{semerakpasj00}.

\section{ BH + Disks}

In this section we  study the superposition of a BH and three
 different models of disks. In the  first case the disk is the 
 relativistic generalization of the Newtonian Kuzmin
 disk that is  studied
 in Ref. \cite{LB}. This solution  is closely related to the vacuum solution 
of the Einstein equation known as  Chazy-Curzon metric \cite{cc}. 
The metric function $\hat\psi$ for the superposition of this
 disk with a BH is,
\be 
\hat\psi(r,0)= -M/(r^2 +a^2)^\half \label{kcc}
\ee
The disk has mass $M$ an infinite radius, $a$ is a parameter related to 
the maximum of the energy density. 

The second and third model of disks are the zeroth and first Morgan and Morgan disks \cite{mm} that are obtained by solving the Laplace equations in oblate coordinates. The function $\hat\psi(r,0)$ for the zeroth order Morgan and Morgan disk is
\ba 
&&\hat\psi(r,0)=-\frac{M}{a}\arctan\frac{a}{\sqrt{r^2 -a^2}}, \;\; r>a  \lb{mm01}\\
&&\hat\psi(r,0)=-\frac{\pi   M}{2a},\;\; r<a .\lb{mm02}
\ea
$M$ represents the mass of the disk and $a$ its radius. This disk has infinite density on the edge. The superposition of this disk and a BH is studied in \cite{letoleq}. 

For the first Morgan and Morgan  disk we have, 
\ba 
&&\hat\psi=-\frac{3 M}{4a^2}[ \sqrt{r^2 -a^2} 
-a^{-1}(r^2 -2a^2)\arctan\frac{a}{\sqrt{r^2 -a^2}} ,\nn\\
&& \hspace{6cm} r>a . \lb{mm11}\\
&&\hat\psi=-\frac{3\pi M}{4a^3}(a^2 -r^2/2), \;\;\;\; r<a \lb{mm12}
\ea
Again $M$ is the mass of the disk and $a$ its radius. The superposition of this disk with a BH was studied in \cite{lemlet}\cite{lemletmm}. This disk has finite density on the rim.
For simplicity we have presented the potentials  Weyl coordinates 
$(t,r,z,\varphi)$.
We note that to  a disk of radius $a$ in Weyl coordinates corresponds  a 
disk of radius 
\be
R_a=m+\sqrt{m^2+a^2}, \lb{Ra}
\ee
 in the usual BH coordinates that will be used in our analysis of stability.
We shall study  numerically the condition
 $hh_{,R}=0$ that will give us information on the stability of circular orbits
for the different superpositions of BH and disks.

\begin{figure} 
\includegraphics[width=2in,height=2.5in,angle=-90]{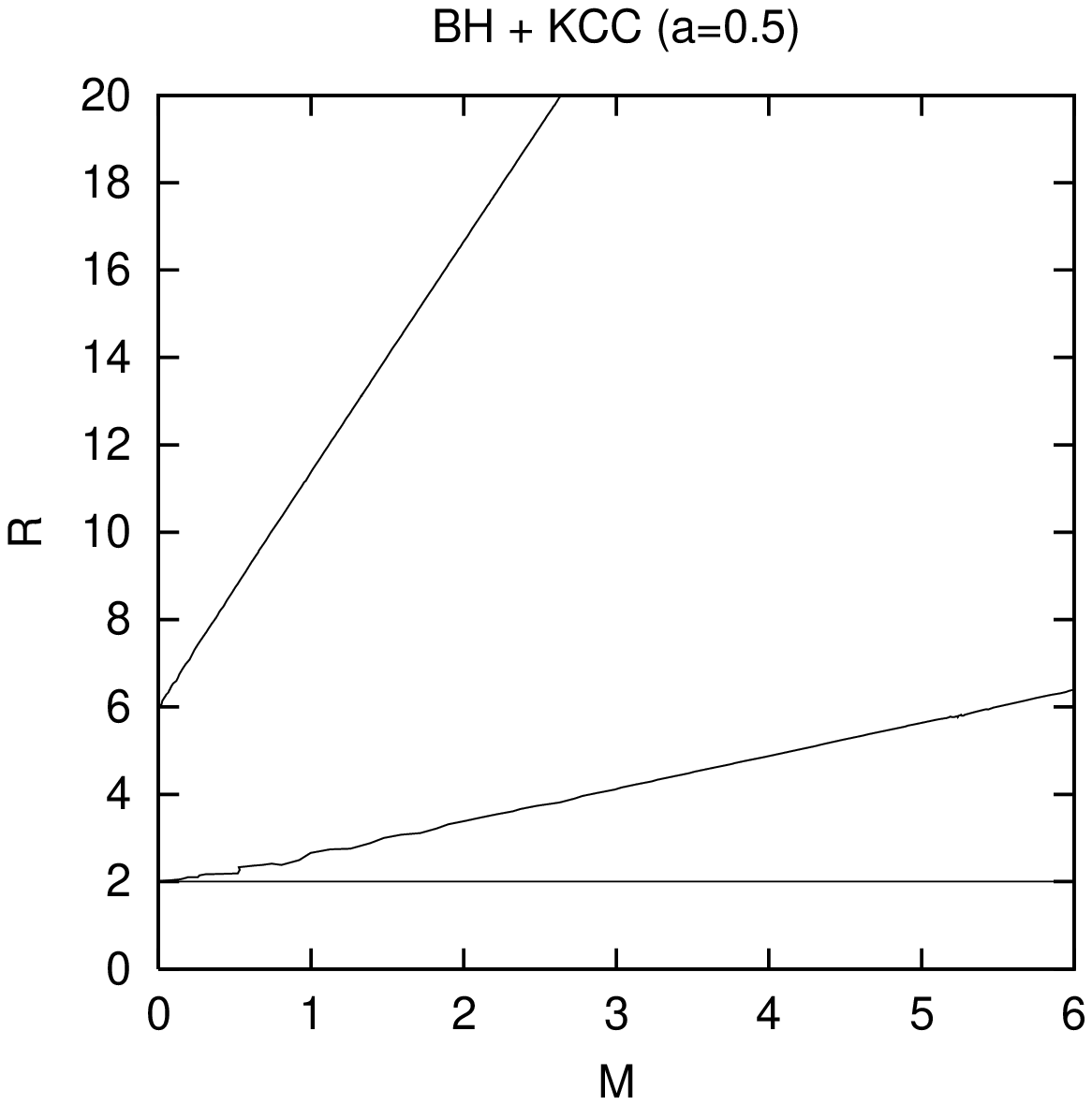}
\includegraphics[width=2in,height=2.5in,angle=-90]{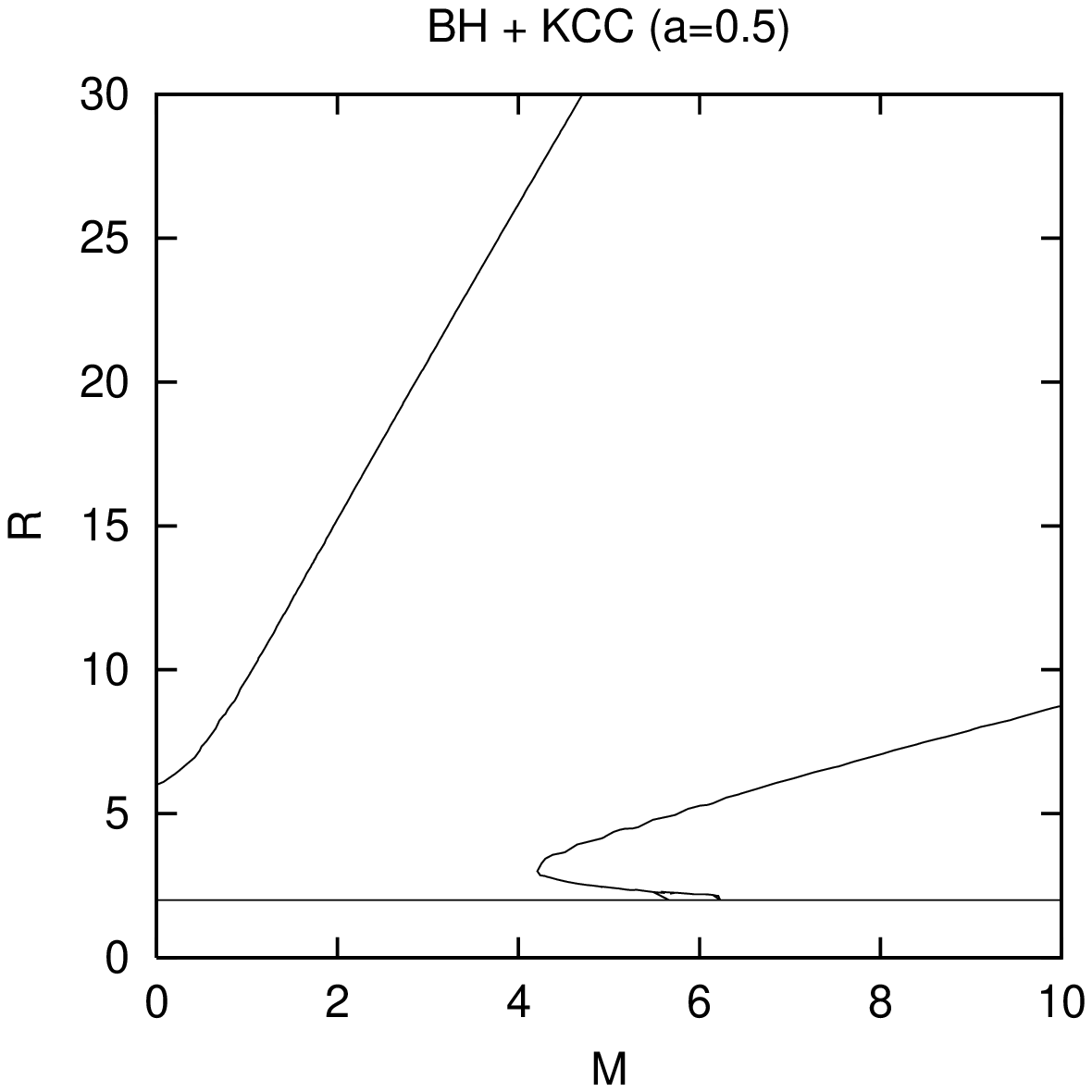} 
\includegraphics[width=2in,height=2.5in,angle=-90]{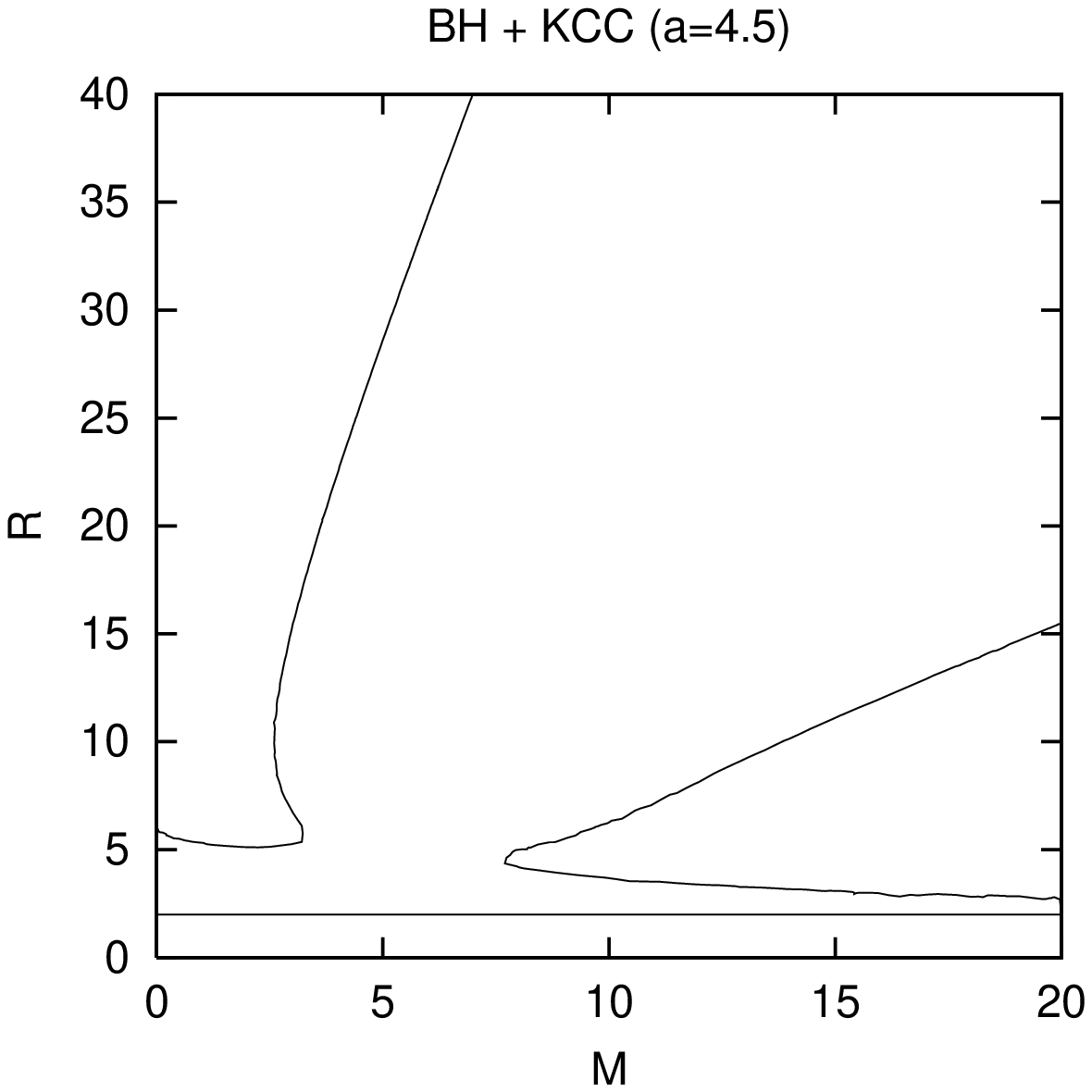}
\includegraphics[width=2in,height=2.5in,angle=-90]{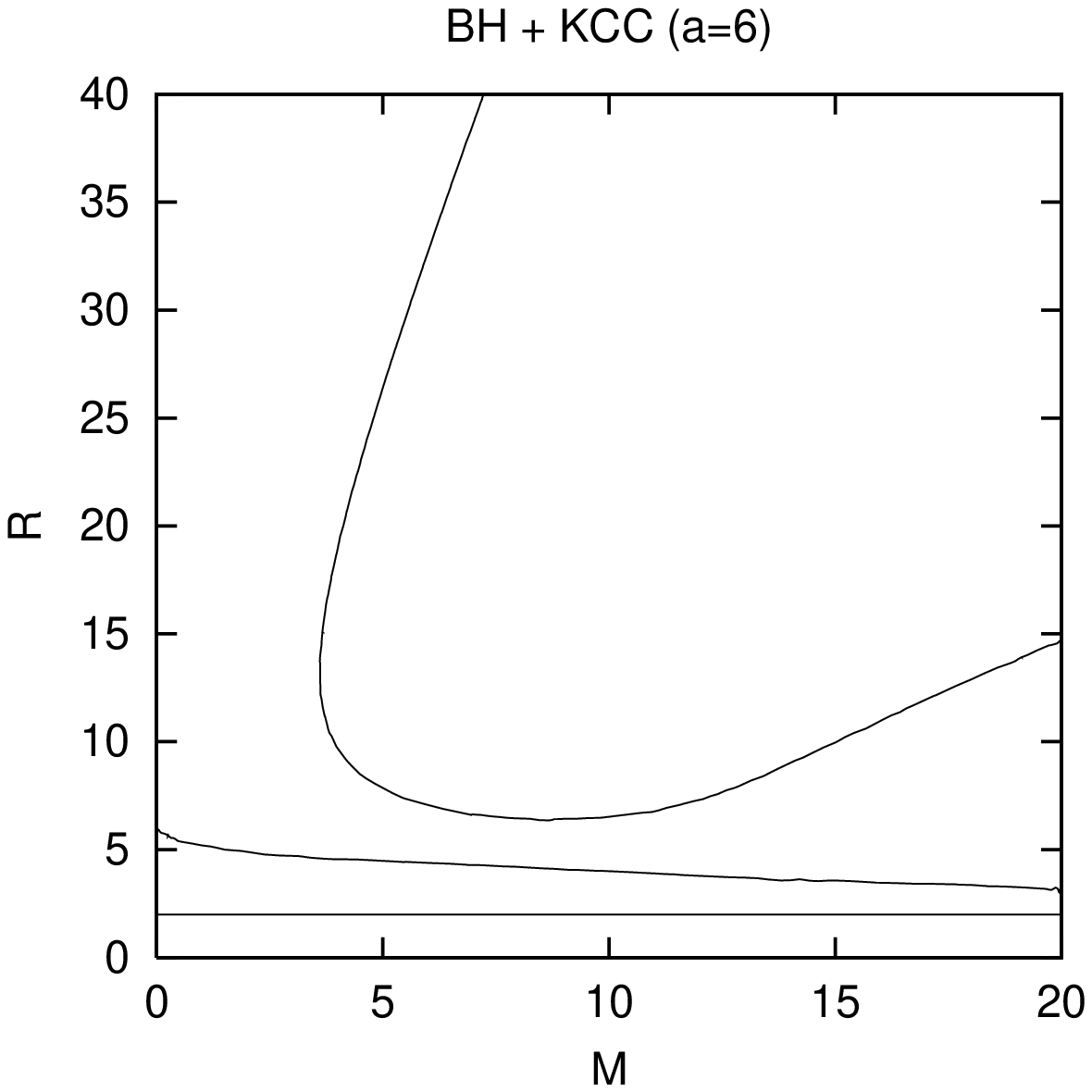}
\caption{ Curves $hh_{,R}=0$ for the superposition of a BH and a
Kuzmin-Chazy-Curzon disk for different values of the parameter $a$. 
 The curve, $R=2m$, is also shown. In the first three cases 
we have stability in
 the regions limited by the upper curves and the axis R and in 
the regions limited by the  lower curves  and  $R=2m$. In the last 
case we have stability in the region limited by the upper curve and the
 lower curve.   We use units such that $G=c=m=1$.                                          }  \label{fig1}
\end{figure}

For the Kuzmin-Chazy-Curzon disk in Fig \ref{fig1}
we study   the curve $hh_{,R}=0$ for different 
values of the parameter $a$, we take $a=0.5,2.5,4.5,6$. Along all this work
we take units such that
$G=c=1$ and we fix the unit of length by taking 
 the mass of the central BH as one.
We have stability in
 the regions limited by the upper curves and the axis R and in 
 regions limited by the  lower curves  and  $R=2m$. In the first case,
$a=0.5$, for a given total mass parameter of the disk, $M$, we have a 
 region, close to the BH horizon, wherein we have  stable circular orbits.
 For   larger values of $R$  we have a  region of instability.
 In principle this region of instability will give rise to  a 
gap in the disk.  For even larger values of
$R$ we will have another region of stability that  extent to infinite. 
In the second case, $a=2.5$,  closer to the horizon, for 
$4<M<6$,  will appear another region of instability. So we can have the formation of stable  flat rings in this region. In the third case, $a=4.5$ we can have rings around $M=3$ and for $M>8$.  In the last case, $a=6$,  rings can also 
be formed.
It is interesting to compare this case with the equivalent Newtonian situation. By applying the RCS to the system represented by the Newtonian potential
\be 
\phi=-M/(R^2+a^2)^\half -m/R \lb{kuzmin}
\ee
we find only  stable circular orbits. Therefore the formations of rings and gaps, in the present case,  is a pure general relativistic effect.

\begin{figure} 
\includegraphics[width=2in,height=2.5in,angle=-90]{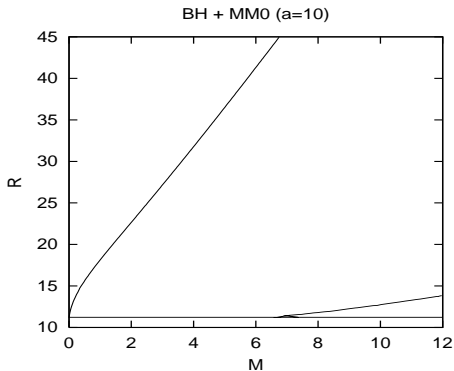}
\caption{ Curves $hh_{,R}=0$ for the superposition of a BH and 
a zeroth order Morgan and Morgan  disk  for $a=10$.
The horizontal line indicates the position of the disk boundary.
We have orbit stability  in
 the region limited by the upper curve and the axis R and in 
the region limited by the  lower curve  and   radius of the disk.
  The units are the same as in the previous figure.  }  \label{fig2}
\end{figure}

Now we shall consider the  zeroth order Morgan and Morgan disk. In the
 interior of the disk we have that the function $\hat\psi$ is constant. Hence
the condition $hh_{,R}=0$ gives us $R=6m$, the LSCO of the usual
  BH. Therefore 
if  the radius of the disk is less than $6m$ it is completely unstable. For 
a disk of radius grater than $6m$ we will have that the region of the disk between $6m$ and its radius will be stable.
The condition  $hh_{,R}=0$ outside the disk is studied in Fig \ref{fig2}.

We have stability in
 the region limited by the upper curve and the axis R and in 
the region limited by the  lower curve  and   the horizontal line that represents the radius of the disk. Therefore we can have
 the formation of a stable structure orbiting around the disk close to its edge. After this zone of stability we find a zone of instability followed by one of stability.

It is instructive to compare these results with the equivalent Newtonian
 situation that is represented by the potential,
\be
\phi=-\frac{M}{a}\arctan\frac{a}{\sqrt{R^2 -a^2}}-\frac{m}{R}. \lb{mm0net}
\ee
We find a very different situation, we only have 
a small zone of instability close to the disk boundary. 
The interior of the disk  is stable.

\begin{figure} 
\includegraphics[width=2in,height=2.5in,angle=-90]{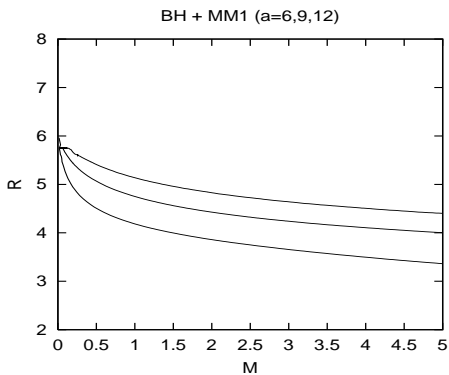}
\caption{ Curves $hh_{,R}=0$ for the superposition of a BH and the first
Morgan and  Morgan disk. The top curve represent a disk with
parameter  $a=12$, the middle with $a=9$ and the bottom curve with $a=6$
We have stable circular orbits in the region  above  each respective curve.
                                        }  \label{fig3}\end{figure}

Now we shall consider the superposition of
first Morgan and Morgan disk with a BH. In Fig \ref{fig3} we
present the curve  $hh_{,R}=0$ for three different disks with parameters $a=12$
(top curve), $a=9$, and $a=6$ (bottom curve). We have a region of instability in the region under the curve  $hh_{,R}=0$ in each case.
We have that the presence of a disk lower the black hole's LSCO.

\begin{figure} 
\includegraphics[width=2in,height=2.5in,angle=-90]{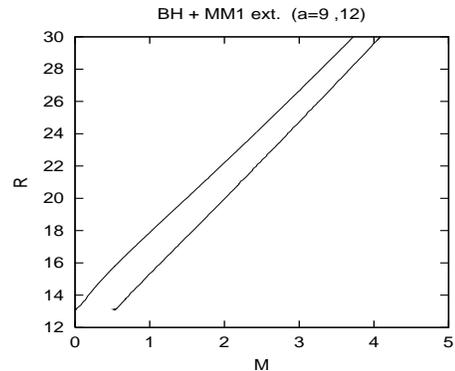}
\caption{ Curves $hh_{,R}=0$ for the superposition of a BH and the first
Morgan and  Morgan disk with $a=9$ (bottom curve) and  $a=12 $ (top curve).
We have stable circular orbits in the region limited by the curve and the axis R. }  \label{fig4}\end{figure}

The  curves $hh_{,R}=0$ for the superposition of a BH and the first Morgan and Morgan disk with $a=9$ is presented in Fig. \ref{fig4}
for a region outside the disk.  We have that the closer to the edge of the disk we always have unstable  circular orbits. For larger radius we have stable orbits. 

It is interesting to compare this case with the same Newtonian situation.
For the interior of the  disk with a  Newtonian center of attraction we find no
regions of instability. Outside the disk the situation is similar to the one shown in Fig. \ref{fig4}, but in  the Newtonian case the curves are lower. 

The   superposition of the BH with the zeroth order Morgan and Morgan 
disk and with the first Morgan and Morgan disk present different properties 
of stability near the edge of the disk. We have that the first mentioned 
disk has a singular rim and the second has a regular edge. This last case
 represent a physically acceptable situation. 

\section{ BH + Rings}

In this section we consider the superposition of a black hole and a family 
of  thin rings.
The function $\hat\psi$  for the  thin rings will be taken as,
\ba 
&&\hat\psi=-\frac{ M}{2}[\frac{1}{\sqrt{r^2 -r^2}}+ \arctan\frac{a}{\sqrt{r^2 -a^2}} ],\;  r>a \lb{r1}\\
&&\hat\psi=-\frac{ M}{2}[ \frac{1}{\sqrt{a^2 -r^2}} +\frac{\pi}{2}],\;\;\;\;  r<a . \lb{r2}
\ea
The parameters $M$ and $a$ represent
 the mass and radius of the ring,  respectively.
This family of thin rings as well as its superposition with
 a black hole   is studied in some detail in \cite{letoleq}.

\begin{figure} 
\includegraphics[width=2in,height=2.5in,angle=-90]{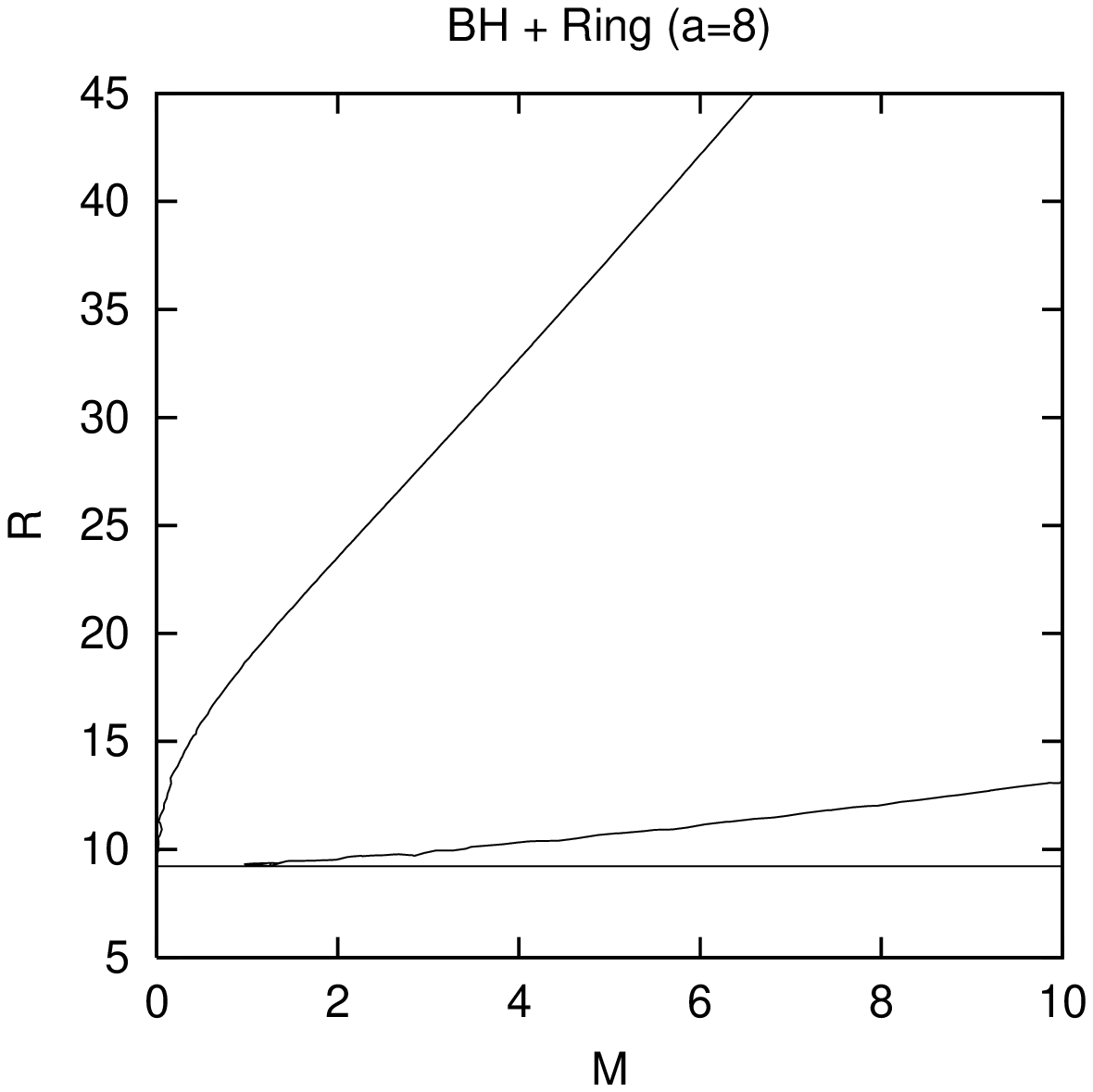}
\includegraphics[width=2in,height=2.5in,angle=-90]{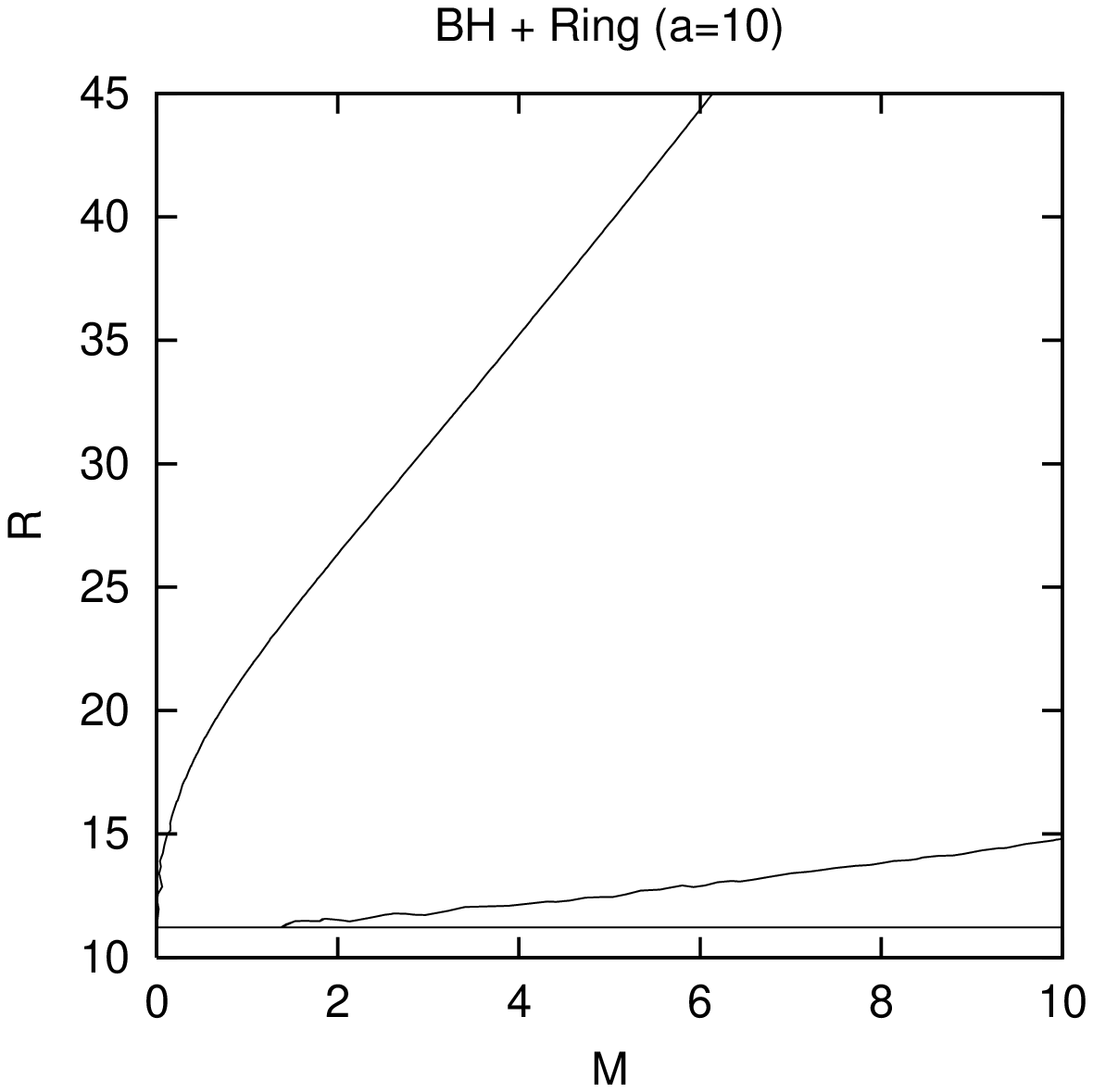}
\includegraphics[width=2in,height=2.5in,angle=-90]{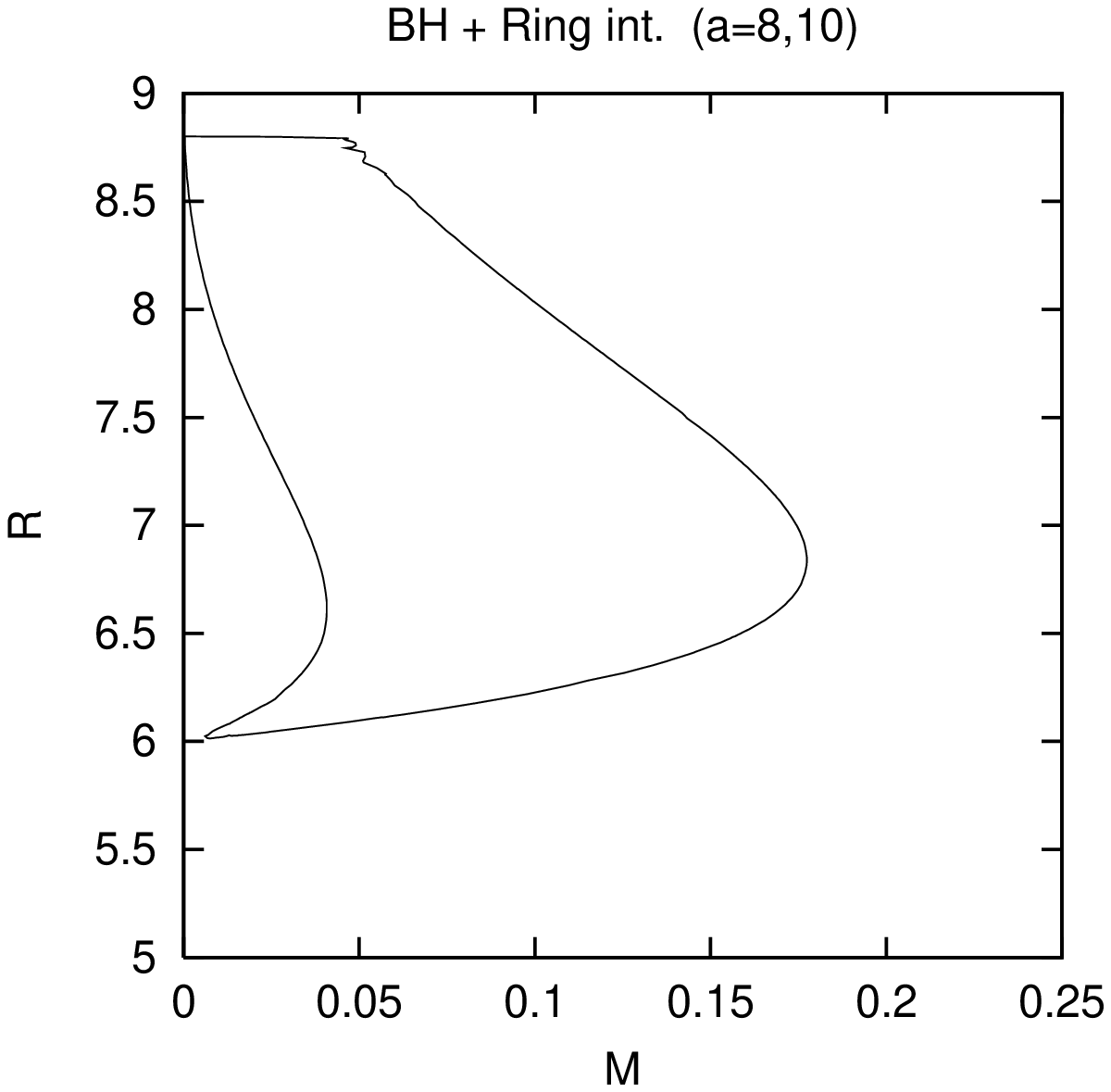}

\caption{ Curves $hh_{,R}=0$ for the superposition of a BH and a thin ring. The first two graphs show the region exterior to  rings with $a=8$ and $a=10$ and the third  the interior region for rings of the same radii  
the LHS (RHS) curve correspond to $a= 8(10)$. In the first two graphs
 we have stable circular orbits in the
 the regions limited by the upper curves and the axis R and in 
the regions limited by the  lower curves  and  the horizontal line that represents the location of the ring. In the third  graph  we have stable circular orbits in the region limited by the corresponding curve and the axis R.
 }  \label{fig5}\end{figure}

In Fig. \ref{fig5} we consider  the curve  $hh_{,R}=0$  for the
 superposition of a BH and the ring with potential $\hat\phi$ given
 by (\ref{r1})-(\ref{r2}).
   The first graph shows the region exterior to the ring and the 
second the interior region for $a=8$ (left hand side curve) and $a=10$ (right hand side curve). For $a=8$($a=10$) we have  $R_8\sim 9$($R_{10}\sim 11$).

 For the exterior region  we have stable circular orbits
for values of $(M,R)$  limited by the upper curve  and the axis R and in 
the region  limited by the  lower curves  and  the horizontal line that represents the location of the ring. In the third graph  we have stable circular orbits in the region limited by the corresponding curve and the axis R.

The regions of stability for the equivalent Newtonian situation are different.
We  have only one small region of instability of circular orbits with 
 radii  larger to the ring radius. For the interior of the ring we have
 that the curve  $hh_{,R}=0$  is similar to the precedent case
      for values of $R$ close  to the ring radius.
 The Newtonian case is similar to the relativistic case, but the curve 
does not turns back to $R=6m$ it decreases as $R\sim 1/M^{1/3}$.
 The analysis of stability of circular orbits for the superposition of a BH and a flat ring (Lemos-Letelier solution \cite{lemlet}), as well as other aspects of this solution were  considered in great detail in \cite{semerakI}, 
\cite{semerakpasj00}, and \cite{semerakIII}.

\section{ BH + Mulipolar fields}

Multipolar exact perturbation  of a BH can approximate all  
situations studied in the previous sections of this article that refer to compact sources of  gravity. Also we can have 
  some insight for the stability of circular orbits in
 the generic case of a static 
BH with   axially
 symmetric perturbations with reflection symmetry.

The functions $\hat\psi(r,z)$ in the present case are
\ba
&&\hat\psi(r,0)=-Q_i/r^3 \lb{qi}\\
&&\hat\psi(r,0)=-Q_e r^2 \lb {qe}
\ea
Where $Q_i$ represents the quadrupolar moment of an internal
 distribution of matter, i.e., of  sources of the gravitational field
  near the BH, e.g.,   disks,  rings, etc.  This field will
  approximate the gravitation field   outside its source.  We have two cases 
$Q_i>0$ and $Q_i<0$. The first case  represents oblate deformations  like 
disks and rings and the  second prolate deformations, e.g.,  jets.
The constant $Q_e$ is related to  the quadrupolar moment of external
 distributions
 of matter, i.e., of sources of the gravitational field  far away
 from the BH, e.g, the gravity in the space limited by  a  ring or
   a shell of matter
 (halo). We also have two different situations  in this
 case $Q_e>0$ and $Q_e<0$. 
The first case describes  oblate deformations like a ring or  oblate 
halos and the second prolate deformations like prolate halos.

\begin{figure} 
\includegraphics[width=2in,height=2.5in,angle=-90]{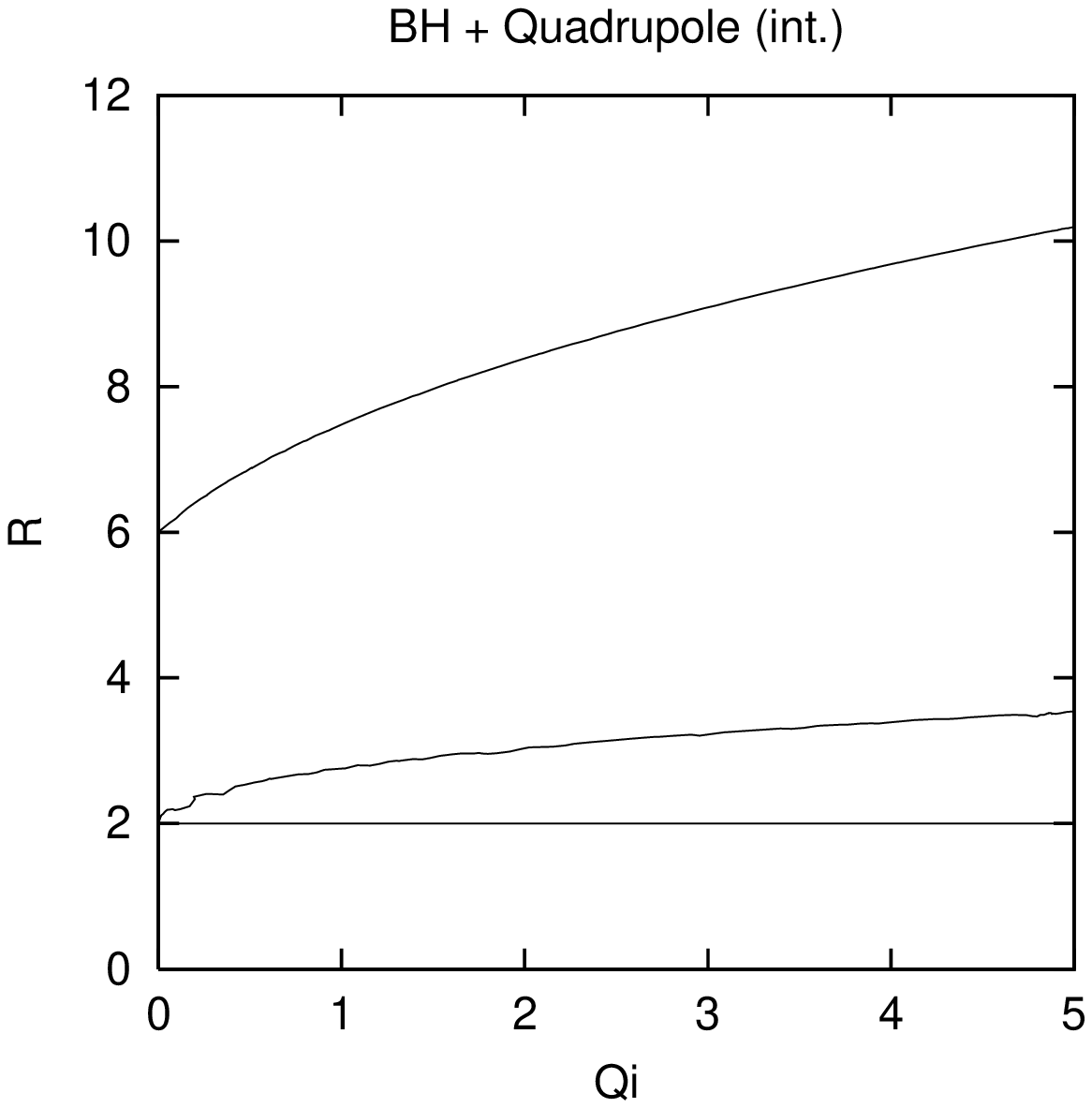}
\includegraphics[width=2in,height=2.5in,angle=-90]{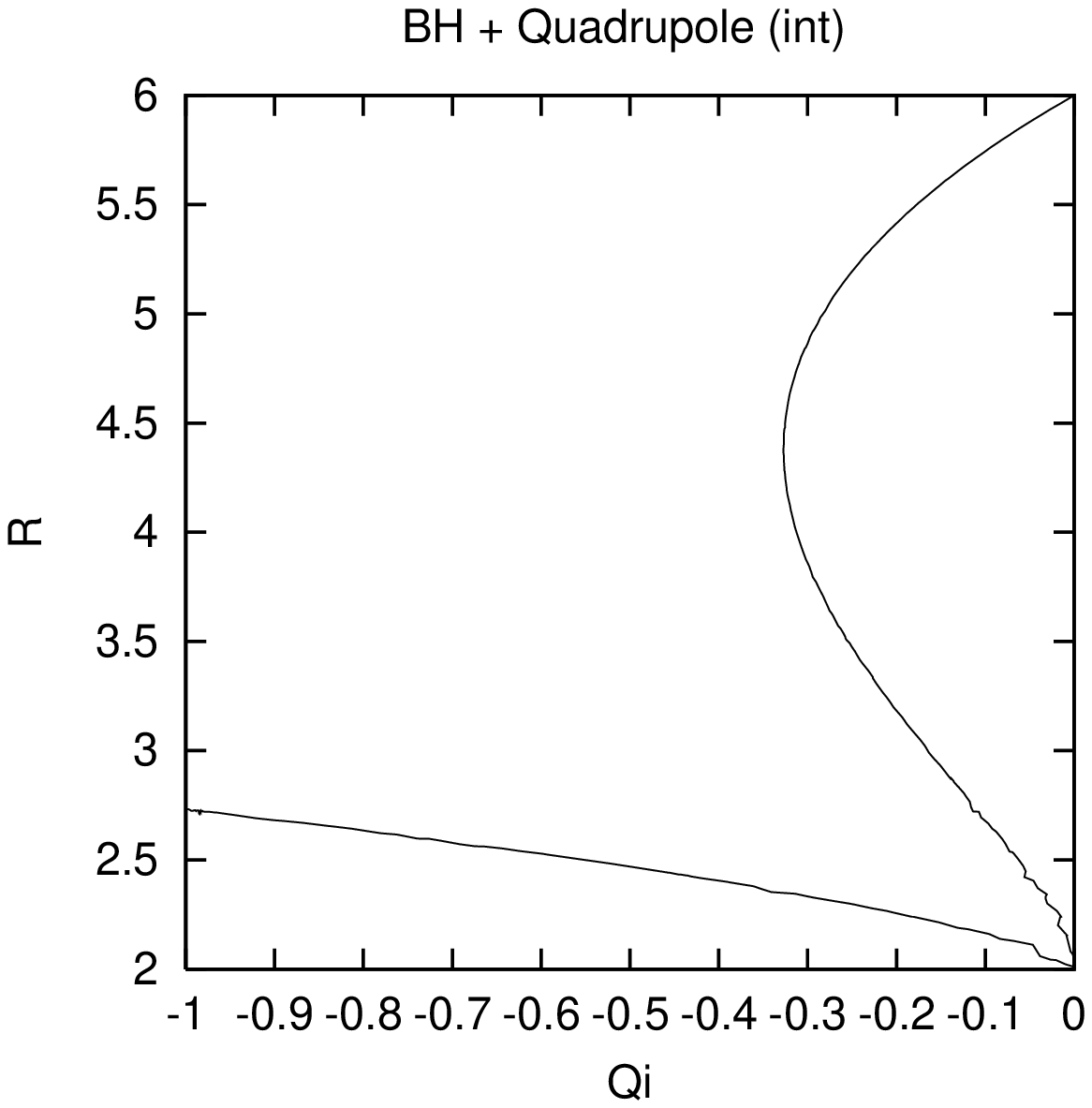}

\caption{ Curves $hh_{,R}=0$ for the superposition of a 
BH and an internal quadrupolar field. 
 The first graph shows  the oblate case and the second 
the prolate deformations.
 In the first we have stable circular orbits in the
 the regions limited by the upper curve  and the axis R and in 
the regions limited by the  lower curves  and  the horizontal 
line that represents the BH horizon. In the second graph  we
 have stable circular orbits in the region limited by
 the axis R and the curves.
 }  \label{fig6}\end{figure}

In Fig. \ref{fig6} we present  curves  $hh_{,R}=0$  for the
 superposition of a BH and an internal multipolar field with
 $\hat\phi$ give by (\ref{qi}). The first graph shows  the oblate case and the second  prolate deformations. In the oblate case 
 we have stable circular orbits in the
 the regions limited by the upper curve  and the axis R and in 
the regions limited by the  lower curves  and  the horizontal line that 
represents the BH horizon. The most relevant region  is the one with $R>6m$. We see
 that the region of stability increases with the internal quadrupole moment. It is interesting to compare this result with the previous case. Let us 
 consider the superposition of a BH with the first Morgan and Morgan disk.
 We see in Fig \ref{fig4} that  for a given disk mass the region of
 instability increases with the  disk radius the same is true for
 rings, see Fig \ref{fig5}. We also checked this property 
for  the zeroth order Morgan and Morgan disk. We have,
  for these compact objects, that  the quadrupolar moments scale as $Q_i\sim M a^2$.  Therefore to  a bigger radius corresponds 
 a bigger quadrupole
 moment  and a larger region of instability. 

In the prolate case, last graph of Fig. \ref{fig6}  we see that
 the quadrupole moment   increases the zone of stability,
with no quadrupole moment we have stability only when $R>6m$. 
This is a rather surprising  effect valid only for equatorial circular 
orbits.  For generic orbits, a prolate internal deformation introduces great instability, moreover we can have  very irregular orbits (chaos) \cite{guelet1}.

In the equivalent Newtonian system, for the oblate case 
we have a region of instability near the center of attraction. The limit curve is $R=(\sqrt{3Q_i/m}$. We have no instability in the prolate  case. Again for
non equatorial circular orbits we can have chaotic motion in this case \cite{guelet1}.

\begin{figure} 
\includegraphics[width=2in,height=2.5in,angle=-90]{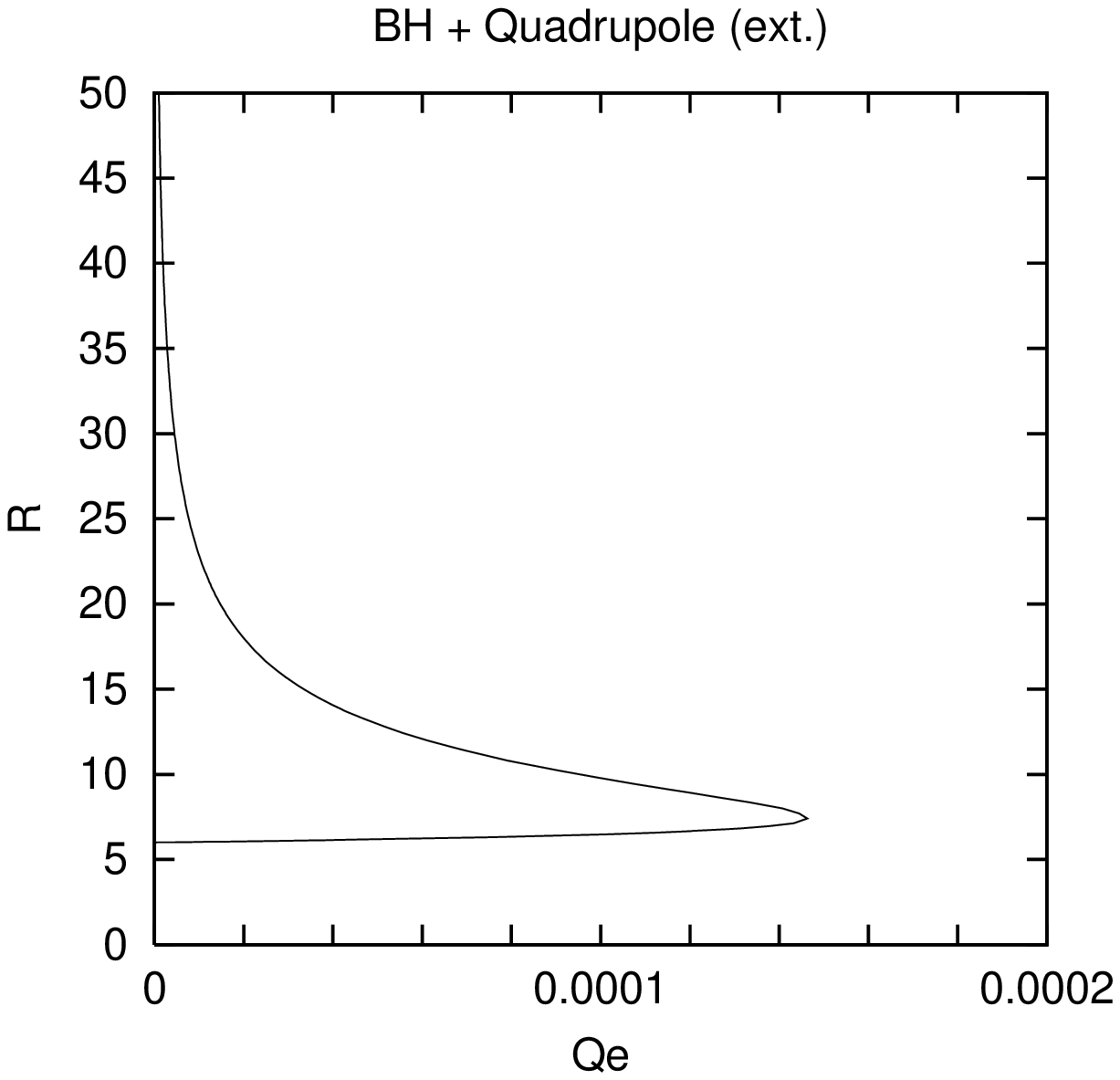}
\includegraphics[width=2in,height=2.5in,angle=-90]{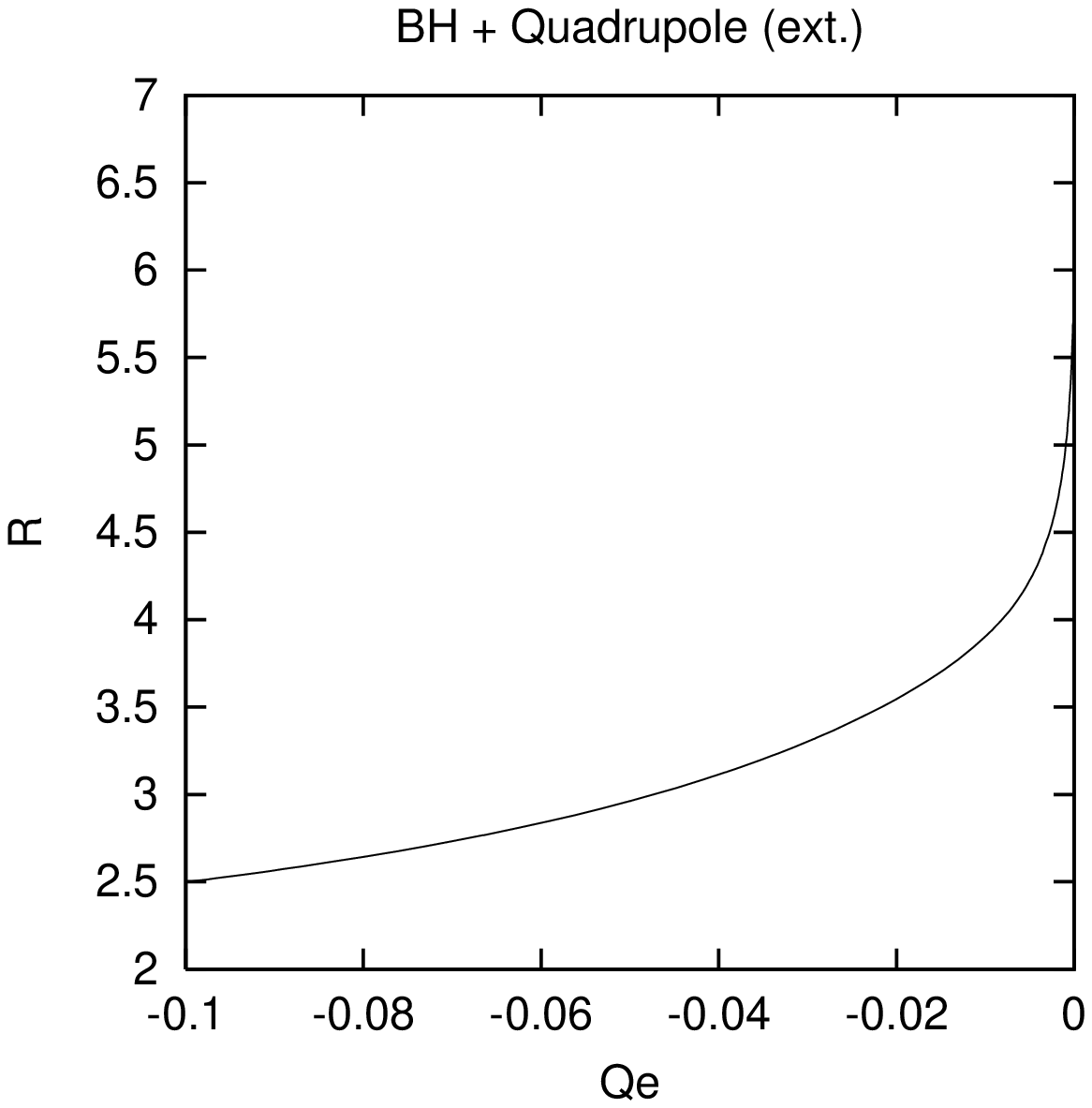}

\caption{ Curves $hh_{,R}=0$ for the superposition of a 
BH and an external quadrupolar field. 
 In the first graph , oblate case,  we have stable circular orbits in the
 the regions limited by the  curve  and the axis R. In the second graph, prolate case,   we have stable circular orbits in the region above the curve.
 }  \label{fig7}\end{figure}

In Fig. \ref{fig7} we present  curves  $hh_{,R}=0$  for the
 superposition of a BH and external multipolar field with
 $\hat\phi$ give by (\ref{qe}). 
In the first graph, oblate case,  we have stable circular orbits in the
 the regions limited by the  curve  and the axis R. We have that this external quadrupole fields severely restrict the possibility of
 stable orbits. We have that for larger value of $Q_e>0.00015$ we do not
 have equatorial stable circular orbits. For 
 $Q_e<0.00015$ we have that the region of stability is limited from above as well as from below. For  generic orbits in BH  exterior halo systems we 
have chaotic orbits \cite{vielet}.

We have that the exterior quadrupole moment of a ring or any external configuration of matter like a shell or halo scales as $Q_e=M/a^3$, with $M$ the mass of the shell and $a$ the typical size.  
Then   large radius indicates  small quadrupole moment and large stability region. We note that second graph of Fig. \ref{fig5} illustrate
  exactly this conclusion.
In the equivalent Newtonian case we have that the curve $hh_{,R}=0$ is
$R=(m/8Q_e)^{1/3}$.

In the prolate case,  second graph of Fig. \ref{fig7},  we have stable
 circular orbits in the region above the curve. We see that  the prolate 
external matter increases the region of stability. For the Newtonian equivalent system we always have stable orbits.
Marginally stable orbits  in quadrupole deformed centers of
 attraction were also studied in \cite{ZGour}.
                                                                                               
\section{Concluding Remarks}

We think  that one of the most  important results shown  in this paper is 
that we can have the formations of rings and gaps around a BH as
  a pure general relativistic effect, of course this is a consequence 
 of the
 presence of the BH horizon that modifies the usual Newtonian 
 effective potential adding
  new critical points.  These critical points change completely the dynamics of the orbiting particles. In particular,
 the maximum  has the property of being a saddle point in phase space. Hence perturbed orbits tend to have irregular motions.

The use of multipoles to represent a distribution of matter in the general
 relativistic context need to be taken with care.  In general relativity, 
the  multipolar expansion take into account that we have a tensorial
 source of the gravitational field and not a scalar one as in
 Newtonian gravity. We
 can have different sets of general relativistic multipoles that in the limit
 of Newtonian gravity  give the same set of 
usual Newtonian  multipoles, for a 
discussion of this point in the context of static axially symmetric spacetimes
 see \cite{bruno}. We found for orbits that are 
 a few Schwarzschild radius away from the center of attraction 
no significant discrepancy  due the use of different multipolar expansion 
with the same Newtonian limit \cite{bruno}.  Hence
 there is no great  error in taken any multipolar
 expansion with the same Newtonian limit, the one taken in this paper
 is the simpler.

  Our study of chaos in BH with halos \cite{vielet} and deformed centers of attractions \cite{guelet1} indicates that the stability of circular orbits is 
quite different from the stability of generic orbits. This is not a 
surprising result since the equatorial circular orbits 
 are a very special case of highly symmetric orbits.

\acknowledgments
I want to thank FAPESP and CNPQ for financial support and Prof. Opher for
getting me interested in the subject of this paper.

\appendix

\section{The counter-rotating hypothesis}
The  EMT for the counter-rotating matter is the sum of the energy momentum tensor of two pressureless streams of particles,
\be
 T^{\mu\nu}=T_+^{\mu\nu}+T_-^{\mu\nu} \lb{c-r}
\ee
where
\be
T_+^{\mu\nu}=\rho_+ U_+^\mu U_+^\nu, \;\; T_- ^{\mu\nu}=\rho_- U_-^\mu U_-^\nu.
\lb{t+t-}
\ee
$\rho_+$ and $\rho_-$ are the matter  density of each stream that are considered equal ($\rho_+ =\rho_-$).  $\;\; U_+^\mu$ and $U_-^\mu$ are the
 corresponding  normalized
 streams'  four-velocities ($U_+^\mu U_{+\mu} =U_-^\mu U_{-\mu} = 1 $).
These four velocities can be written as 
\be
U_+^\mu=\alpha U^\mu+\beta \phi^\nu,\;\; U_-^\mu=\alpha U^\mu-\beta \phi^\nu,
\lb{u+-}
\ee
where $U^\mu$ is a timelike vector and $\phi^\nu$ an spacelike vector tangent to one of the two streams of particles, also $U^\mu U_\mu=-\phi^\nu\phi_\nu=1,
\;\; U^\mu\phi_\nu=0$. Note that  $\alpha^2-\beta^2=1$. Therefore the EMT 
(\ref{c-r}) can we put in the form
 \be 
T^{\mu\nu}=\rho U^\mu U^\nu +p_\phi \phi^\mu\phi^\nu,\lb{tmn}
\ee
with $\rho=2\alpha^2\rho_+$ and $p_\phi=2(1-\alpha^2)\rho$.

Thus the counter-rotating hypothesis is consistent with the disks represented
 by the  Weyl metrics solutions to the Einstein  equations that have only 
azimuthal pressure. Furthermore we can 
assume that
\be 
\nabla_\mu T_+^{\mu\nu}=\nabla_\mu T_-^{\mu\nu}=0. \lb{bianchi1}
\ee
From (\ref{t+t-}) we have $U_+^\mu\nabla_\mu U_+^\nu =
U_-^\mu\nabla_\mu  U_-^\nu=0$, i.e., each stream follows a geodesic motion.
Note that  (\ref{bianchi1}) 
  implies that $\nabla_\mu T^{\mu\nu}=0$. Thus the supposition that
 each stream follows a geodesic flow is consistent with the Bianchi identity for
 the Weyl metric. For  multifluids, see for instance Ref. \cite{multf}.

In summary, when the counter-rotating  hypothesis is assumed, 
the stability of circular geodesics is equivalent to the stability of the flow of each fluid component. 
For  the counter-rotating model of cold disks we have that the
 LSCO defines  exactly the inner radius of the disk.


\begin{thebibliography}{99}

\bb{Lord} Lord Rayleigh,  Proc. R. Soc. Lond.
 Ser. A 93, 148 (1916). See also, L. D. Landau, E. M. Lifshitz, ``Fluid Mechanics'', 2nd Ed. (Pergamon Press, Oxford, 1987), \S 27.

\bb{LB}
J. Bi\u{c}\'{a}k, D. Lynden-Bell and J. Katz, Phys. Rev.
D47, 4334 (1993) 

\bb{hot}  P. S. Letelier. Phys. Rev. D 60, 104042-1 (1999).

\bb{voghtletdisk} D. Voght and P.S. Letelier, Phys. Rev. D, to appear.
Available at http://xxx.lanl.gov/abs/gr-qc/0308031.

\bb{bardeen} J.M. Bardeen, Ap. J., 161, 103 (1970).

\bb{seguin}  F.H. Seguin, Ap. J. 197, 745 (1975).

\bb{marek}  M.A. Abramowicz, Mon. Not. R. astr. Soc. 245, 720 (1990).

\bb{W} See for instance, H. Robertson and T. Noonan, 
``Relativity and Cosmology"(Saunders, London 1968) pp  272-278. 

\bibitem{counter} F. Bertola {\it et al.}
Ap. J.  458, L67 (1996).

\bb{mm} T. Morgan and L. Morgan, Phys. Rev. 183, 1097 (1969).

\bb{semerakpasj00} O. Semer\'ak and M. \v Z\'a\v cek, Publ. Astron. Soc. Japan, 52, 1067 (2000) 


\bb{cc} M. Chazy, Bull. Soc. Math. France 52, 17 (1924); H. Curzon, Proc. London Math. Soc. 23, 477 (1924).


\bb{letoleq}  P.S.  Letelier and S.R. Oliveira ,  Class. Quantum  Grav. 15, 
421 (1998)
\bb{lemlet}J.P.S. Lemos and P.S. Letelier, Phys. Rev. D 49, 5135 (1994).

\bb{lemletmm} J.P. Lemos and P.S.  Letelier,  Class. Quantum  Grav. 10, 
L75 (1993)
 
\bb{semerakI} O. Semer\'ak and M. \v Z\'a\v cek, Class. Quantum Grav. 17, 1613 (2000).

 \bb{semerakIII} O. Semer\'ak, Class. Quantum Grav. 20, 1613 (2003).

\bb{guelet1} E. Gu\'eron  and P.S.  Letelier, Phys.  Rev. E 66, 046611  (2002);
{\em ibid}  63, 035201(R), (2001).

\bb{vielet}
W.M. Vieira and P.S. Letelier,  Ap.
 J.  513, 383 (1999).

\bb{ZGour} J.L. Zdunik and E. Gourgoulhon, Phys. Rev. D 63, 087501  (2001).

\bb{bruno} B. Boisseau and P.S. Letelier,  Gen.  Rel. Grav. 34, 1077 (2002)



\bb{multf} P.S.  Letelier,  Phys.  Rev.  D 22, 807 (1980)


\end{thebibliography}
\end{document}